\documentstyle[12pt,epsf]{article}

  \def\pp{{\mathchoice
          {
              \kern 1pt%
              \raise 1pt
              \vbox{\hrule width5pt height0.4pt depth0pt
                    \kern -2pt
                    \hbox{\kern 2.3pt
                          \vrule width0.4pt height6pt depth0pt
                          }
                    \kern -2pt
                    \hrule width5pt height0.4pt depth0pt}%
                    \kern 1pt
           }
            {
              \kern 1pt%
              \raise 1pt
              \vbox{\hrule width4.3pt height0.4pt depth0pt
                    \kern -1.8pt
                    \hbox{\kern 1.95pt
                          \vrule width0.4pt height5.4pt depth0pt
                          }
                    \kern -1.8pt
                    \hrule width4.3pt height0.4pt depth0pt}%
                    \kern 1pt
            }
            {
              \kern 0.5pt%
              \raise 1pt
              \vbox{\hrule width4.0pt height0.3pt depth0pt
                    \kern -1.9pt  
                    \hbox{\kern 1.85pt
                          \vrule width0.3pt height5.7pt depth0pt
                          }
                    \kern -1.9pt
                    \hrule width4.0pt height0.3pt depth0pt}%
                    \kern 0.5pt
            }
            {
              \kern 0.5pt%
              \raise 1pt
              \vbox{\hrule width3.6pt height0.3pt depth0pt
                    \kern -1.5pt
                    \hbox{\kern 1.65pt
                          \vrule width0.3pt height4.5pt depth0pt
                          }
                    \kern -1.5pt
                    \hrule width3.6pt height0.3pt depth0pt}%
                    \kern 0.5pt
            }
        }}

  \def\mm{{\mathchoice
                       {
                             \kern 1pt
               \raise 1pt    \vbox{\hrule width5pt height0.4pt depth0pt
                                  \kern 2pt
                                  \hrule width5pt height0.4pt depth0pt}
                             \kern 1pt}
                       {
                            \kern 1pt
               \raise 1pt \vbox{\hrule width4.3pt height0.4pt depth0pt
                                  \kern 1.8pt
                                  \hrule width4.3pt height0.4pt depth0pt}
                             \kern 1pt}
                       {
                            \kern 0.5pt
               \raise 1pt
                            \vbox{\hrule width4.0pt height0.3pt depth0pt
                                  \kern 1.9pt
                                  \hrule width4.0pt height0.3pt depth0pt}
                            \kern 1pt}
                       {
                           \kern 0.5pt
             \raise 1pt  \vbox{\hrule width3.6pt height0.3pt depth0pt
                                  \kern 1.5pt
                                  \hrule width3.6pt height0.3pt depth0pt}
                           \kern 0.5pt}
                       }}

\catcode`@=11
\def\un#1{\relax\ifmmode\@@underline#1\ese
        $\@@underline{\hbox{#1}}$\relax\fi}
\catcode`@=12

\let\du=\du                     

\def\a{\alpha}
\def\b{\beta}

\def\d{\delta}

\def\f{\phi}

\def\l{\lambda}
\def\m{\mu}
\def\n{\nu}

\def\q{\theta}

\def\s{\sigma}

\def\ve{\varepsilon}

\def\vq{\vartheta}

\def\cd{{\cal D}}

\def\bo{{\raise-.5ex\hbox{\large$\Box$}}}               
\def\pa{\partial}                                       
\def\TH{{\raise.2ex\hbox{$\displaystyle \bigodot$}\mskip-4.7mu \llap H \;}}
\def\face{{\raise.2ex\hbox{$\displaystyle \bigodot$}\mskip-2.2mu \llap {$\ddot
        \smile$}}}                                      

\def\Bar#1{\overline{#1}}                       
\def\leftrightarrowfill{$\mathsurround=0pt \mathord\leftarrow \mkern-6mu
        \cleaders\hbox{$\mkern-2mu \mathord- \mkern-2mu$}\hfill
        \mkern-6mu \mathord\rightarrow$}
\def\dvec#1{\vbox{\ialign{##\crcr
        \leftrightarrowfill\crcr\noalign{\kern-1pt\nointerlineskip}
        $\hfil\displaystyle{#1}\hfil$\crcr}}}           
\def\dt#1{{\buildrel {\hbox{\LARGE .}} \over {#1}}}     

\def\frac#1#2{{\textstyle{#1\over\vphantom2\smash{\raise.20ex
        \hbox{$\scriptstyle{#2}$}}}}}                   
\def\sfrac#1#2{{\vphantom1\smash{\lower.5ex\hbox{\small$#1$}}\over
        \vphantom1\smash{\raise.4ex\hbox{\small$#2$}}}} 
\def\bfrac#1#2{{\vphantom1\smash{\lower.5ex\hbox{$#1$}}\over
        \vphantom1\smash{\raise.3ex\hbox{$#2$}}}}       
\def\afrac#1#2{{\vphantom1\smash{\lower.5ex\hbox{$#1$}}\over#2}}    
\def\on#1#2{\mathop{\null#2}\limits^{#1}}               
\def\bvec#1{\on\leftarrow{#1}}                  

\def\[{\lfloor{\hskip 0.35pt}\!\!\!\lceil}
\def\]{\rfloor{\hskip 0.35pt}\!\!\!\rceil}
\def\Lag{{\cal L}}
\def\du#1#2{_{#1}{}^{#2}}

\def\fracm#1#2{\hbox{\large{${\frac{{#1}}{{#2}}}$}}}
\def\ha{{\fracmm12}}

\def\Tr{{\rm Tr}}

\def\un{\underline}
\def\fracmm#1#2{{{#1}\over{#2}}}

\def\low#1{{\raise -3pt\hbox{${\hskip 0.75pt}\!_{#1}$}}}

\def\Dot#1{\buildrel{_{_{\hskip 0.01in}\bullet}}\over{#1}}
\def\dt#1{\Dot{#1}}

\newskip\humongous \humongous=0pt plus 1000pt minus 1000pt
\def\caja{\mathsurround=0pt}
\def\eqalign#1{\,\vcenter{\openup2\jot \caja
        \ialign{\strut \hfil$\displaystyle{##}$&$
        \displaystyle{{}##}$\hfil\crcr#1\crcr}}\,}
\newif\ifdtup

\parskip=\medskipamount                 
\thicklines                         

\begin{document}
\thispagestyle{empty}

{\hbox to\hsize{
\vbox{\noindent July 2004 \hfill hep-th/0407211 }}}

\noindent
\vskip1.3cm
\begin{center}

{\Large\bf $SU(2)\times U(1)$ Non-anticommutative N=2 
                   Supersymmetric Gauge Theory 
~\footnote{Supported in part by the JSPS and the Volkswagen Stiftung}}
\vglue.2in

Sergei V. Ketov~\footnote{Email address: ketov@phys.metro-u.ac.jp}
and Shin Sasaki~\footnote{Email address: shin-s@phys.metro-u.ac.jp}

{\it Department of Physics, Faculty of Science\\
          Tokyo Metropolitan University\\
         1--1 Minami-osawa, Hachioji-shi\\
            Tokyo 192--0397, Japan}
\end{center}

\vglue.2in

\begin{center}
{\Large\bf Abstract}
\end{center}

\noindent
We derive the master function governing the component action of the 
four-dimensional {\it non-anticommutative} (NAC) and fully N=2 supersymmetric 
gauge field theory with a non-simple gauge group $U(2)=SU(2)\times U(1)$. We 
use a Lorentz-singlet NAC-deformation parameter and an N=2 supersymmetric star
(Moyal) product, which do not break any of the fundamental symmetries of the 
undeformed N=2 gauge theory. The scalar potential in the NAC-deformed theory 
is calculated. We also propose the non-abelian BPS-type equations in the case 
of the NAC-deformed N=2 gauge theory with the $SU(2)$ gauge group, and comment 
on the $SU(3)$ case too. The NAC-deformed field theories can be thought of as 
the effective (non-perturbative) N=2 gauge field theories in a certain (scalar
 only) N=2 supergravity background. 

\newpage

\section{Introduction}

Noncommutative spaces were extensively studied in the past. The simplest and 
best known example of noncommutativity is provided by the phase space 
coordinates in quantum mechanics. In quantum field theory, both at the 
perturbative and non-perturbative level, the assumption of spacetime 
noncommutativity is known to lead to new physical phenomena, such as UV/IR 
mixing, noncommutative solitons, quantum Hall fluid, etc. (see e.g., 
refs.~\cite{rev,li} for a review and an extensive list of references). The 
spacetime noncommutativity introduces non-locality into field theory in a mild
 and controllable manner. In particular, a noncommutative field theory still 
possesses a chiral ring, and there exists a change of variables (the so-called
 Seiberg-Witten map) that brings the gauge transformations to the standard 
form \cite{sw,sei}.

Supersymmetric gauge field theories in {\it Non-AntiCommutative} (NAC) 
superspace \cite{nac} is a rather new area of research \cite{rec}. Those
NAC-deformed field theories naturally arise from superstrings in certain 
supergravity backgrounds, being natural extensions of the usual (undeformed) 
supersymmetric gauge field theories. In string theory the noncommutativity of 
bosonic spacetime coordinates naturally emerges in a (multiple) D-brane 
worldvolume, when a constant NS--NS two-form is turned on \cite{sw}. More 
recently, in the context of the Dijkgraaf-Vafa correspondence \cite{dv} 
relating N=1 supersymmetric gauge theories and matrix models, it was suggested
 \cite{ovn} that the non-anticommutativity of superspace coordinates naturally 
appears in a D-brane worldvolume when a constant RR two-form is turned on in 
ten dimensions (see also ref.~\cite{ll}). A similar phenomenon was discovered 
in four dimensions when a constant self-dual graviphoton field strength is 
taken as a superstring background \cite{sei}.

Fermionic non-anticommutativity means that the odd superspace coordinates
obey a Clifford algebra instead of being anticommuting \cite{nac}. It is also
possible to keep the commutativity of the bosonic spacetime coordinates,
which renders the NAC-deformed field theory much more tractable \cite{sei}.
Consistency implies that merely a chiral part of the fermionic superspace 
coordinates should become NAC, whereas the anti-chiral fermionic superspace 
coordinates should be kept anticommuting (in some basis). This is only 
possible when the anti-chiral fermionic coordinates $(\bar{\q})$ are {\it not}
 complex conjugates to the chiral ones, $\bar{\q}\neq(\q)^*$, which is the 
case in Euclidean and Atiyah-Ward spacetimes with the signature $(4,0)$ and 
$(2,2)$, respectively. The Euclidean signature is relevant to instantons and 
superstrings \cite{rev,rec}, whereas the Atiyah-Ward signature is relevant to 
the critical N=2 string models \cite{ov} and the supersymmetric self-dual
 gauge field theories \cite{mary}.

Extended supersymmetry offers more opportunities depending upon how much of
supersymmetry one wants to preserve, as well as which NAC deformation (e.g., 
a singlet or a non-singlet) and which operators (the supercovariant derivatives
or the supersymmetry generators) one wants to employ in the Moyal-Weyl star
product \cite{nac,2nac}. The $N=(1,1)$ (or just $N=2$) extended supersymmetry 
is very special in that respect since it allows one to choose a singlet NAC 
deformation and a star product that preserve all the fundamental symmetries 
\cite{fs}. Indeed, the most general nilpotent deformation of 
$N=(1,1)=2\times(\ha,\ha)$ supersymmetry is given by 
$$ \{ \q_i^{\a},\q_j^{\b}\}_{\star}=\d^{(\a\b)}_{(ij)}C^{(\a\b)} -
2iP\ve^{\a\b}\ve_{ij}\quad {\rm (no~sum!)}~,\eqno(1.1)$$
where $\a,\b=1,2$ are chiral spinor indices, $i,j=1,2$ are the indices of the 
internal R-symmetry group $SU(2)_R$, while $C^{\a\b}$ and $P$ are some 
constants. Taking only a singlet deformation to be non-vanishing, $P\neq 0$, 
and using the chiral supercovariant N=2 superspace derivatives  $D_{i\a}$ in 
the Moyal-Weyl star product,
$$ A\star B = A\exp\left( iP\ve^{\a\b}\ve^{ij}\bvec{D}{}_{i\a}\vec{D}{}_{j\b}
\right)B~~,\eqno(1.2)$$ 
allows one to keep manifest N=2 supersymmetry, Lorentz invariance and 
R-invariance, as well as (undeformed) gauge invariance (after some non-linear
field redefinition) \cite{fs}. The star product (1.2) matching those
conditions is unique, and it requires $N=2$.

We choose flat Euclidean spacetime for definiteness, but continue to use the 
notation common to N=2 superspace with Minkowski spacetime signature, as it is
 becoming increasingly customary in the current literature 
(see also ref.~\cite{book} for more details about our notation). Our NAC N=2 
superspace with the coordinates $(x^m,\q^{i}_{\a},\bar{\q}_{i}^{\dt{\a}})$ is 
defined by eq.~(1.1), with $C^{\a\b}=0$ and $P\neq 0$, as the only non-trivial 
(anti)commutator amongst the N=2 superspace coordinates. This choice preserves
all most fundamental symmetries and features of N=2 supersymmetry including the
so-called G-analyticity \cite{fs}.

A NAC-deformed (non-abelian) supersymmetric gauge field theory can also be 
rewritten to the usual form, with the standard gauge transformations of its 
field components, i.e. as some kind of effective action, after certain 
(non-linear) field redefinition known as the Seiberg-Witten map 
({\it cf.} ref.~\cite{sw}). In the case of the $P$-deformed N=2
 super-Yang-Mills theory such (non-abelian) map was calculated by 
Ferrara and Sokatchev in ref.~\cite{fs} with the following result for the 
effective anti-chiral N=2 superfield strength:
$$  \Bar{W}_{\rm NAC}= \fracmm{\Bar{W}}{1+P\Bar{W}}~~.\eqno(1.3)$$
Here $\Bar{W}$ is the standard (Lie algebra-valued) covariant N=2 superfield 
strength subject to the standard N=2 superspace Bianchi identities
$$ \cd_{i\a}\Bar{W}=0 \qquad {\rm and}\qquad
\cd_{ij}W=\Bar{\cd}_{ij}\Bar{W}~~,\eqno(1.4)$$
in terms of the N=2 superspace gauge- and super-covariant derivatives 
$\cd^{i\a}$ and $\Bar{\cd}_{i\dt{\a}}$, obeying an algebra
$$ \{ \cd^i_{\a}, \cd^j_{\b} \}= -2\ve^{ij}\ve_{\a\b}\Bar{W}~.\eqno(1.5)$$ 
We use the notation $\cd\low{ij}=\cd^{\a}\low{(i}\cd\low{j)\a}$ and 
$\Bar{\cd}\low{ij}=\Bar{\cd}_{\dt{\a}(i}\Bar{\cd}^{\dt{\a}}\low{j)}$, and define 
the covariant field components of the N=2 superfield $\Bar{W}$ by covariant 
differentiation,
$$ \Bar{W}|=\bar{\f}~,\quad \Bar{\cd}_{i\dt{\a}}\Bar{W}|=
\bar{\l}_{i\dt{\a}}~,\quad \Bar{\cd}_{ij} \Bar{W}|=D_{ij}~,\quad
\Bar{\cd}_{\dt{\a}\dt{\b}} \Bar{W}|=F_{\dt{\a}\dt{\b}}~,\eqno(1.6)$$
where $|$ denotes the leading ($\q$- and $\bar{\q}$-independent) component of
an N=2 superfield.

The effective N=2 superspace action reads
$$ S_{\rm NAC}=\fracm{1}{2}\int d^4x\low{R} 
d^4\bar{\q}\,\Tr\, \Bar{W}^2_{\rm NAC} \equiv \fracm{1}{2} \int d^4x\low{R}
d^4\bar{\q}\,\Tr f(\Bar{W})~, \eqno(1.7)$$
whose structure function $f(\Bar{W})$ follows from eq.~(1.3), 
$$ f(\Bar{W})= \left(\fracmm{\Bar{W}}{1+P\Bar{W}}\right)^2~~.\eqno(1.8)$$

It is non-trivial to calculate the action (1.7) in components because of the 
need to perform the (non-abelian) group-theoretical trace (the Lagrangian is 
no longer quadratic in $\Bar{W}\,$!). The case of the NAC, N=2 supersymmetric 
gauge field theory with an (abelian) $U(1)$ gauge group is, of course, fully
straightforward because its master function, governing the action of its field
components after taking the group-theoretical trace, is still given by the 
same function (1.8) --- see e.g., ref.~\cite{fs}. The full equations of motion
 in the NAC-deformed abelian N=2 theory, as well as their BPS-like 
counterparts, were calculated in our earlier paper \cite{ks1}. The master 
function in the simplest non-abelian, NAC and  N=2 supersymmetric gauge field 
theory with the gauge group $SU(2)$ was found in ref.~\cite{ks2}. In this 
paper we calculate the master function of the NAC four-dimensional N=2 
supersymmetric gauge field theory having a non-simple gauge group 
$U(2)=SU(2)\times U(1)$. Our new solution interpolates  between the master 
functions found in refs.~\cite{fs,ks1} and \cite{ks2}.

Our paper is organized as follows. In sect.~2 we perform the $U(2)$ 
group-theoretical trace in eq.~(1.7) in order to find the master function of 
the colorless variables $\Bar{W}^a\Bar{W}^a$ and $\Bar{W}{}^0$ associated with 
the $SU(2)$ and $U(1)$ factors, respectively, which governs the full component
 action. In sect.~3 we show how our results reduce to the known master 
functions for the $SU(2)$ and $U(1)$ gauge groups, separately \cite{ks1,ks2}. 
We also give some new results about the BPS equations in the (non-abelian) 
$SU(2)$ case. In sect.~4 we calculate the scalar potential in the 
deformed $U(2)$ theory. Sect.~5 is our conclusion that includes a short 
discussion of the $SU(3)$ case too.

\section{Calculation of the $U(2)$ trace}

In the $U(2)$ case we find convenient to use the hermitian $3\times 3$
matrices~\footnote{The anti-hermitian generators in the case of the $SU(2)$ 
gauge group were used in ref.~\cite{ks2}.} both for the $U(1)$ and the 
$SU(2)$ generators, namely,
$$T^0 = \left( \begin{array}{ccc} 1 & 0 & 0 \\ 0 & 1 & 0 \\ 0 & 0 & 1
\end{array}\right) \eqno(2.1a)$$
and
$$ T^1 = \left( \begin{array}{ccc} 0 & 0 & 0 \\ 0 & 0 & -i \\ 0 & i & 0
\end{array}\right)~~,\quad
T^2 = \left( \begin{array}{ccc} 0 & 0 & i \\ 0 & 0 & 0 \\ -i & 0 & 0
\end{array}\right)~~,\quad
T^3 = \left( \begin{array}{ccc} 0 & -i & 0 \\ i & 0 & 0 \\ 0 & 0 & 0
\end{array}\right)~~,\eqno(2.1b)$$
respectively, which obey the commutation relations $\[T^a,T^b\]=i\ve^{abc}T^c$,
where $\ve^{abc}$ is the totally antisymmetric Levi-Civita symbol with 
$\ve^{123}=1$ and $a,b,\ldots=1,2,3$.

The master function in the $U(2)$ case under investigation is given by
$$\eqalign{
h(\Bar{W}{}^0,\Bar{W}{}^a) & \equiv \Tr\left[ f(\Bar{W}{}^aT^a+\Bar{W}{}^0T^0)
\right]  = \Tr \left[ \fracmm{\Bar{W}{}^aT^a+\Bar{W}{}^0T^0}{1+
P(\Bar{W}{}^aT^a+\Bar{W}{}^0T^0)}\right]^2 \cr
& = \sum^{+\infty}_{n=0}(n+1)(-)^nP^n\Tr\,G^{n+2}~,\cr}\eqno(2.2)$$
where we have introduced the notation
$$ G =\Bar{W}{}^aT^a+\Bar{W}{}^0T^0~,\eqno(2.3)$$
and $\Tr$ stands for the group-theoretical trace. In particular, we have
$$ \Tr\,G^{n}= \Tr\left(\Bar{W}{}^aT^a+\Bar{W}{}^0T^0 \right)^n=
\Tr\sum^{n}_{r=0}
\left( \begin{array}{c} n \\ r\end{array}\right) 
(\Bar{W}{}^aT^a)^{n-r}(\Bar{W}{}^0T^0)^r~~,\eqno(2.4)$$
where 
$$ \left( \begin{array}{c} n \\ r\end{array}\right) = \fracmm{n!}{r!(n-r)!}~.
\eqno(2.5)$$
The basic $SU(2)$ traces were already computed in ref.~\cite{ks2},
$$\Tr(\Bar{W}{}^aT^a)^{2m}=2(\Bar{W}{}^a\Bar{W}{}^a)^m~,~~m>0,\quad
\Tr(\Bar{W}{}^aT^a)^{0}=3~,\quad \Tr(\Bar{W}{}^aT^a)^{2m-1}=0~,\eqno(2.6)$$ 
so that it is useful to compute the sums over the even and odd powers of $P$ 
in eq.~(2.2) separately. As regards the sum over all even powers of $P$ on the 
right-hand-side of eq.~(2.2), we find
$$ \sum^{\infty}_{m=1}\sum^{m}_{r=0} (2m-1)P^{2m-2} \left(
\begin{array}{c} 2m \\ 2r\end{array}\right) (\Bar{W}{}^0)^{2r}
\Tr(\Bar{W}{}^aT^a)^{2(m-r)}
=  \sum^{\infty}_{m=1}\fracmm{2m-1}{P^2}(P^2\Bar{W}{}^0\Bar{W}{}^0)^m $$
$$ + \sum^{\infty}_{m=1}\fracmm{2m-1}{P^2}\left[ P^2 \Bar{W}{}^0\Bar{W}{}^0
\left( 1-\sqrt{ \fracmm{\Bar{W}{}^0\Bar{W}{}^0}{\Bar{W}{}^a\Bar{W}{}^a}}
\right)^2\right]^m $$
$$ + \sum^{\infty}_{m=1}\fracmm{2m-1}{P^2}\left[ P^2 \Bar{W}{}^0\Bar{W}{}^0
\left( 1+\sqrt{ \fracmm{\Bar{W}{}^0\Bar{W}{}^0}{\Bar{W}{}^a\Bar{W}{}^a}}
\right)^2\right]^m~~,\eqno(2.7)$$
where we have used the identity
$$ \sum^{m-1}_{r=0}\left( \begin{array}{c} 2m \\ 2r\end{array}\right)x^r=
\fracm{1}{2}\left[(1-\sqrt{x})^{2m}+(1+\sqrt{x})^{2m}-2x^m\right]~.\eqno(2.8)$$
Next, when using the identities
$$ \sum^{\infty}_{m=1}x^m =\fracmm{x}{1-x}\quad {\rm and}\quad
 \sum^{\infty}_{m=1}mx^m =\fracmm{x}{(1-x)^2}~~~~,\eqno(2.9)$$
we can rewrite eq.~(2.7) to the form
$$\eqalign{
&~ \fracmm{\Bar{W}{}^0\Bar{W}{}^0(1+P^2\Bar{W}{}^0\Bar{W}{}^0)}{(1-P^2
\Bar{W}{}^0\Bar{W}{}^0)^2} ~~+\cr
 +~~ &~ \fracmm{1+P^2\Bar{W}{}^a\Bar{W}{}^a \left(1-
\sqrt{ \fracmm{\Bar{W}{}^0\Bar{W}{}^0}{\Bar{W}{}^a\Bar{W}{}^a}}\right)^2}{
\left[ 1-P^2\Bar{W}{}^a\Bar{W}{}^a\left(1-
\sqrt{\fracmm{\Bar{W}{}^0\Bar{W}{}^0}{\Bar{W}{}^a\Bar{W}{}^a}}\right)^2
\right]^2}\Bar{W}{}^a\Bar{W}{}^a
\left(1-\sqrt{\fracmm{\Bar{W}{}^0\Bar{W}{}^0}{\Bar{W}{}^a\Bar{W}{}^a}}
\right)^2 ~~+ \cr
 +~~ &~ \fracmm{1+P^2\Bar{W}{}^a\Bar{W}{}^a \left(1+
\sqrt{ \fracmm{\Bar{W}{}^0\Bar{W}{}^0}{\Bar{W}{}^a\Bar{W}{}^a}}\right)^2}{
\left[ 1+P^2\Bar{W}{}^a\Bar{W}{}^a\left(1+
\sqrt{\fracmm{\Bar{W}{}^0\Bar{W}{}^0}{\Bar{W}{}^a\Bar{W}{}^a}}\right)^2
\right]^2}\Bar{W}{}^a\Bar{W}{}^a
\left(1+\sqrt{\fracmm{\Bar{W}{}^0\Bar{W}{}^0}{\Bar{W}{}^a\Bar{W}{}^a}}
\right)^2~~~~.\cr}\eqno(2.10)$$

Similarly, the sum over odd powers of $P$ on the right-hand-side of 
eq.~(2.2) is given by
$$  \sum^{\infty}_{m=1}\sum^{m}_{r=0} 2mP^{2m-1} \left(
\begin{array}{c} 2m+1 \\ 2r+1\end{array}\right) (\Bar{W}{}^0)^{2r+1}
\Tr(\Bar{W}{}^aT^a)^{2(m-r)}
= \fracmm{2P(\Bar{W}{}^0)^3}{(1-P^2\Bar{W}{}^0\Bar{W}{}^0)^2} ~~+$$
$$+~ \left( 1+ \fracmm{1}{\sqrt{\fracmm{\Bar{W}{}^0\Bar{W}{}^0}{\Bar{W}{}^a
\Bar{W}{}^a}}}\right)
 \fracmm{2P\Bar{W}{}^0(\Bar{W}{}^a\Bar{W}{}^a)\left(1+
\sqrt{ \fracmm{\Bar{W}{}^0\Bar{W}{}^0}{\Bar{W}{}^a\Bar{W}{}^a}}\right)^2}{
\left[ 1-P^2(\Bar{W}{}^a\Bar{W}{}^a)\left(1+
\sqrt{\fracmm{\Bar{W}{}^0\Bar{W}{}^0}{\Bar{W}{}^a\Bar{W}{}^a}}\right)^2
\right]^2} ~~+ \eqno(2.11)$$
$$ +~ \left( 1- \fracmm{1}{\sqrt{\fracmm{\Bar{W}{}^0\Bar{W}{}^0}{\Bar{W}{}^a
\Bar{W}{}^a}}}\right)
 \fracmm{2P\Bar{W}{}^0(\Bar{W}{}^a\Bar{W}{}^a)\left(1-
\sqrt{ \fracmm{\Bar{W}{}^0\Bar{W}{}^0}{\Bar{W}{}^a\Bar{W}{}^a}}\right)^2}{
\left[ 1-P^2(\Bar{W}{}^a\Bar{W}{}^a)\left(1-
\sqrt{\fracmm{\Bar{W}{}^0\Bar{W}{}^0}{\Bar{W}{}^a\Bar{W}{}^a}}\right)^2
\right]^2}~~~~,$$
where we have used yet another identity
$$ \eqalign{
\sum^{m-1}_{r=0}\left( \begin{array}{c} 2m+1 \\ 2r+1\end{array}\right)x^r =~ & 
\fracmm{1}{2}\left[(1+\sqrt{x})^{2m}+(1-\sqrt{x})^{2m}\right] ~+ \cr
& +\fracmm{1}{2\sqrt{x}}\left[(1+\sqrt{x})^{2m}-(1-\sqrt{x})^{2m}\right]-x^m~,
\cr}\eqno(2.12)$$
together with eqs.~(2.6) and (2.9). The final result for the $U(2)$ master
function is given by a sum of eqs.~(2.10) and (2.11), which reads
$$\eqalign{
h(\Bar{W}{}^0,\Bar{W}{}^a)  = & \left( \fracmm{\Bar{W}{}^0}{1+P\Bar{W}{}^0}
\right)^2 ~+ \cr
& + \left[ \fracmm{\Bar{W}{}^0+\sqrt{\Bar{W}{}^a\Bar{W}{}^a}}{1+P\left(
\Bar{W}{}^0+\sqrt{\Bar{W}{}^a\Bar{W}{}^a}\right)}\right]^2 +
\left[ \fracmm{\Bar{W}{}^0-\sqrt{\Bar{W}{}^a\Bar{W}{}^a}}{1+P\left(
\Bar{W}{}^0-\sqrt{\Bar{W}{}^a\Bar{W}{}^a}\right)}\right]^2~~.\cr}\eqno(2.13)$$
This equation is one of the main new results of our paper, because it is needed
for a straightforward calculation of the component action out of eqs.~(1.6) 
and (1.7).

\section{Some limits, and non-abelian BPS equations}

In this section we are going to demonstrate how some earlier established 
results \cite{fs,ks1,ks2} follow from our general equation (2.13), as well as
find new (non-abelian) BPS equations in the NAC, N=2 theory with the $SU(2)$ 
gauge group.

(i) Firstly, as regards the commutative limit $P\to 0$, we easily find
$$\eqalign{
\lim_{P\to 0} h(\Bar{W}{}^0,\Bar{W}{}^a) &~ = \left(\Bar{W}{}^0\right)^2 +
\left( \Bar{W}{}^0 + \sqrt{\Bar{W}{}^a\Bar{W}{}^a}\right)^2 +
\left( \Bar{W}{}^0 - \sqrt{\Bar{W}{}^a\Bar{W}{}^a}\right)^2  \cr 
&~ = 3(\Bar{W}{}^0)^2 + 2\Bar{W}{}^a\Bar{W}{}^a=\Tr(\Bar{W}{}^aT^a+
\Bar{W}{}^0T^0)^2~~.\cr}\eqno(3.1)$$
This reproduces the usual (commutative) N=2 supersymmetric $U(2)$ gauge theory,
as it should.

(ii) Second, let us consider another limit, $\Bar{W}{}^0\to 0$. In this case 
eq.~(2.13) yields
$$\eqalign{
\lim_{\Bar{W}{}^0\to 0} h(\Bar{W}{}^0,\Bar{W}{}^a) &~ = 
\left( \fracmm{ \sqrt{\Bar{W}{}^a \Bar{W}{}^a}}{1+
P\sqrt{\Bar{W}{}^a\Bar{W}{}^a}}\right)^2 +
\left( \fracmm{ -\sqrt{\Bar{W}{}^a \Bar{W}{}^a}}{1-
P\sqrt{\Bar{W}{}^a\Bar{W}{}^a}}\right)^2 \cr
&~ = \fracmm{2\Bar{W}{}^a\Bar{W}{}^a +2P^2(\Bar{W}{}^a\Bar{W}{}^a)^2}{(1-P^2
 \Bar{W}{}^a\Bar{W}{}^a)^2}~~.\cr}\eqno(3.2)$$
This result precisely reproduces the master function in the NAC, N=2 
supersymmetric Yang-Mills theory with the gauge group $SU(2)$, which was 
calculated in ref.~\cite{ks2}. 

(iii) Third, in the NAC abelian limit $\Bar{W}{}^a\to 0$, we find from 
eq.~(2.13) that 
$$\eqalign{
\lim_{\Bar{W}{}^a\to 0} h(\Bar{W}{}^0,\Bar{W}{}^a) &~ =
\left( \fracmm{\Bar{W}{}^0}{1+P\Bar{W}{}^0}\right)^2
+2\left( \fracmm{\Bar{W}{}^0}{1+P\Bar{W}{}^0}\right)^2 \cr
&~ =3\left(\fracmm{\Bar{W}{}^0}{1+P\Bar{W}{}^0}\right)^2
= \Tr\left(\fracmm{\Bar{W}{}^0{\bf 1}}{{\bf 1}+P\Bar{W}{}^0{\bf 1}}\right)^2 
~~,\cr}\eqno(3.3)$$
where ${\bf 1}=T^0$ stands for a unit $3\times 3$ matrix. Equation (3.3)  
precisely reproduces the NAC, N=2 supersymmetric gauge theory with the abelian
 gauge group $U(1)$ \cite{fs,ks1}.

The BPS equations in the non-anticommutative N=2 supersymmetric gauge theory 
with the {\it abelian} gauge group $U(1)$ were derived in ref.~\cite{ks1}. To 
this end, we would like to derive the {\it non-abelian} BPS equations, by 
considering the non-anticommutative N=2 gauge theory with a simple gauge
group $SU(2)$ for simplicity. The component Lagrangian of this theory was
calculated in ref.~\cite{ks1} by decomposing the relevant N=2 
anti-chiral superfield $\Bar{W}{}^2\equiv\Bar{W}{}^a\Bar{W}{}^a$ as follows 
(in the anti-chiral N=2 basis):
$$\Bar{W}{}^2 = U +V_{\dt{\a}i}\bar{\q}^{\dt{\a}i}+X_{ij}\bar{\q}^{ij}
+Y_{\dt{\a}\dt{\b}}\bar{\q}^{\dt{\a}\dt{\b}}
+Z_{\dt{\a}i}(\bar{\q}^3)^{\dt{\a}i} + L\bar{\q}^4~,\eqno(3.4)$$
where we have introduced its (composite) field components 
$(U,V_{\dt{\a}i},X_{ij},Y_{\dt{\a}\dt{\b}},Z_{\dt{\a}i},L)$. The composites
can be expressed in terms of the gauge-covariant field components (1.6) as
follows \cite{ks2}:
$$U=\bar{\f}^a\bar{\f}^a,~~
V_{\dt{\a}i}=2\bar{\l}^a_{\dt{\a}i}\bar{\f}^a,~~
X_{ij}= 2\left(\bar{\f}^aD^a_{ij}-\bar{\l}^{\dt{\a}a}_i
\bar{\l}^{a}_{j\dt{\a}}\right)~,~~
Y_{\dt{\a}\dt{\b}}  =2\left(\bar{\f}^aF^a_{\dt{\a}\dt{\b}}-
\bar{\l}^{ia}_{\dt{\a}}\bar{\l}^{a}_{i\dt{\b}}\right)~,$$
$$Z_{i\dt{\a}} = 4i\bar{\f}^a(\tilde{\s}^{\m})\du{\dt{\a}}{\a}
\cd_{\m}\l^a_{i\a} + \bar{\l}^{ja}_{\dt{\a}}D_{ij} 
- \bar{\l}^{\dt{\b}a}_{i}F^a_{\dt{\a}\dt{\b}}~~,\eqno(3.5a)$$
and
$$\eqalign{
 L  = ~&~ -2\bar{\f}^a\cd_{\m}\cd^{\m}\f^a -i\bar{\l}^a_{i\dt{\a}}
(\tilde{\s}^{\m})^{\dt{\a}\a}\cd_{\m}\l^{ia}_{\a} 
 + \ve^{abc}\l^{ia\a}\bar{\f}^b\l^c_{i\a} +
\ve^{abc}\bar{\l}^{a}_{i\dt{\a}}\f^b\bar{\l}^{\dt{\a}ic} \cr
~&~ + \fracmm{1}{48}D^a_{ij}D^{aij}- \fracmm{1}{12}F^{\m\n a-}F^{a-}_{\m\n} 
 -\fracm{1}{2}\f^a\bar{\f}^b\f^c\bar{\f}^d\ve^{abf}\ve^{cdf}~,
\cr}\eqno(3.5b)$$
where $\cd_{\m}$ are the usual gauge-covariant derivatives (in the adjoint), 
$F^-_{\m\n}$ is the anti-self-dual part of $F_{\m\n}$, with $\m,\n=1,2,3,4$.
The last composite field $L$ is nothing but the usual (undeformed) 
N=2 super-Yang-Mills Lagrangian,  $L=\Lag_{\rm SYM}$.

The full component action was calculated out of eqs.~(1.7) and (3.2) 
in ref.~\cite{ks1},
$$\eqalign{ 
\Lag_{\rm deformed~SYM} = ~& F(\bar{\f}^2)\Lag_{\rm SYM}
+2F'(\bar{\f}^2) \left[ -4i\bar{\f}^a\bar{\f}^b
(\bar{\l}^a_i\tilde{\s}^{\m}\cd_{\m}\l^{ib}) + 
\bar{\f}^a(\bar{\l}^2)^{ab}_{ij}D^{ijb} \right. \cr
~& +8\bar{\f}^aD^a_{ij}(\bar{\l}^2)^{ij}
 +4\bar{\f}^a\bar{\f}^bD^a_{ij}D^{bij} 
 -\bar{\f}^a(\bar{\l}^2)^{ab}_{\dt{\a}\dt{\b}}
(\tilde{\s}^{\m\n})^{\dt{\a}\dt{\b}}F^{b-}_{\m\n} \cr
~& \left.- 8\bar{\f}^aF^{a-}_{\m\n}(\tilde{\s}^{\m\n})_{\dt{\a}\dt{\b}}
(\bar{\l}^2)^{\dt{\a}\dt{\b}}-128\bar{\f}^a\bar{\f}^bF^{a-}_{\m\n}F^{\m\n b-}
+ 96 \bar{\l}^4 \right] \cr
~&  +8F''(\bar{\f}^2) \left[ -\bar{\f}^a\bar{\f}^b\bar{\f}^c
(\bar{\l}^2)^{ab}_{\dt{\a}\dt{\b}}(\tilde{\s}^{\m\n})^{\dt{\a}\dt{\b}}
F^{c-}_{\m\n}  + \bar{\f}^a\bar{\f}^b\bar{\f}^c(\bar{\l}^2)^{ab}_{ij}D^{ijc}
\right. \cr
~& \left. + 24 \bar{\f}^a\bar{\f}^b(\bar{\l}^4)^{ab}\right] 
 + 16 F'''(\bar{\f}^2) \bar{\f}^a\bar{\f}^b\bar{\f}^c\bar{\f}^d
(\bar{\l}^4)^{abcd}~,\cr}\eqno(3.6)$$  
where we have introduced the book-keeping notation~\footnote{The parameter
$\tilde{P}=iP$ here coincides with the parameter $P$ used in ref.~\cite{ks2}.} 
$$ F(\bar{\f}^2) = \fracmm{1-3\tilde{P}^2\bar{\f}^2}{(1+
\tilde{P}^2\bar{\f}^2)^3}= 1 -6\tilde{P}^2\bar{\f}^2+{\cal O}(\tilde{P}^4
\bar{\f}^4)~,\eqno(3.7)$$
as well as 
$$\eqalign{
 (\bar{\l}^2)_{ij} & 
=\bar{\l}^a_{i\dt{\a}}\bar{\l}_j^{\dt{\a}a} \quad {\rm and}
\quad (\bar{\l}^2)_{\dt{\a}\dt{\b}}=
\bar{\l}^a_{i\dt{\a}}\bar{\l}^{ia}_{\dt{\b}}~,\cr
(\bar{\l}^2)_{ij}^{ab} & =\bar{\l}^a_{i\dt{\a}}\bar{\l}_j^{\dt{\a}b} 
\quad {\rm and} \quad 
(\bar{\l}^2)^{ab}_{\dt{\a}\dt{\b}}
=\bar{\l}^a_{i\dt{\a}}\bar{\l}^{ib}_{\dt{\b}}~~,\cr
(\bar{\l}^2)^{ab} & = \bar{\l}^a_{i\dt{\a}}\bar{\l}^{i\dt{\a}b}
 \quad {\rm and} \quad 
(\bar{\l}^4)^{abcd}  = \bar{\l}^{1a}_{\dt{1}}\bar{\l}^{1b}_{\dt{2}}  
\bar{\l}^{2c}_{\dt{1}} \bar{\l}^{2d}_{\dt{2}}~,\cr}\eqno(3.8)$$
while the primes in eq.~(3.6) denote differentiations with respect to the
argument $\Bar{\f}^2$. The Euler-Lagrange equations of motion of the theory 
(3.6) can be found in  ref.~\cite{ks2}.

To illustrate our procedure of deducing BPS equations from a given action,
let us first consider only the gauge field-dependent terms in eq.~(3.6),
$$ L_{\rm gauge}= -g^{ab}F^{a-}_{\m\n}F^{\m\n b-} +
B^a_{\m\n}F^{\m\n a-}~,\eqno(3.9)$$
where we have introduced the field-dependent `metric' $g^{ab}=g^{ba}$,
$$ g^{ab}(\bar{\f})=\fracm{1}{12}F(\bar{\f}^2)\d^{ab}+2^8F'(\bar{\f}^2)
\bar{\f}^{a}\bar{\f}^b~~,\eqno(3.10)$$
and the antisymmetric (field-dependent) `tensor' $B^a_{\m\n}=-B^a_{\n\m}$,
$$ B^a_{\m\n}(\bar{\f},\bar{\l})=-2(\tilde{\s}_{\m\n})^{\dt{\a}\dt{\b}}\left[
16F(\bar{\f}^2)\bar{\f}^a\bar{\l}_{\dt{\a}\dt{\b}}+F'(\bar{\f}^2)\bar{\f}^b
\bar{\l}^{ba}_{\dt{\a}\dt{\b}}+4F''(\bar{\f}^2)\bar{\f}^a\bar{\f}^b\bar{\f}^c
\bar{\l}^{bc}_{\dt{\a}\dt{\b}}\right]~.\eqno(3.11)$$
                                                             
The most naive BPS procedure amounts to forming perfect squares out of the 
various terms in the Lagrangian, and demanding each squared term to vanish, 
thus `minimizing' the Euclidean action. Applying this procedure to the terms
(3.9) gives rise to the BPS equations
$$ g^{ab}(\bar{\f})F^{b-}_{\m\n}=\fracm{1}{2}B^a_{\m\n}~,\eqno(3.12)$$ 
which non-trivially generalize the non-abelian anti-self-duality condition 
$F^{a-}_{\m\n}=0$. It is worth mentioning that the antisymmetric tensor 
$B^c_{\m\n}(\bar{\f},\bar{\l})$ defined by eq.~(3.11) is automatically 
anti-self-dual. Strictly speaking, we should also demand that the `metric' is
 positively definite, which implies a restriction
$$ g^{ab}(\bar{\f}) >  0~~.\eqno(3.13)$$

Similarly, or just by using the Euler-Lagrange equations of motion, we obtain
from eq.~(3.6) the equations on the auxiliary fields 
$$ h^{ab}(\bar{\f})D^{b}_{ij}+\fracm{1}{2}C^a_{ij}=0~,\eqno(3.14)$$  
where
$$ h^{ab}(\bar{\f})=\fracm{1}{48}F(\bar{\f}^2)\d^{ab}+8F'(\bar{\f}^2)
\bar{\f}^a\bar{\f}^b \eqno(3.15)$$
and
$$ C^{ija}(\bar{\f},\bar{\l})=2F'(\bar{\f}^2)\bar{\f}^b(\bar{\l}^{ij})^{ba}
+16F'(\bar{\f}^2)\bar{\f}^a\bar{\l}^{ij}+
8F''(\bar{\f}^2)\bar{\f}^a\bar{\f}^b\bar{\f}^c(\bar{\l}^{ij})^{bc}~.
\eqno(3.16)$$
Clearly, the algebraic equations (3.14) for $D^a_{ij}$ can be easily solved
as long as $\det h\neq 0$.

The BPS equations have to imply the Euler-Lagrange equations of motion. For 
instance, we checked it  in the NAC, $U(1)$-based N=2 theory in 
ref.~\cite{ks1}. Here, in the $SU(2)$ case, we confine ourselves to the purely
bosonic terms only, for simplicity. Of course, the BPS equation (3.12) gets
undeformed in such situation. However, it is worth mentioning that the NAC 
deformation gives rise to the new, purely bosonic terms in eq.~(3.6). 
Therefore, checking the equations of motion is still non-trivial even when all 
the fermionic fields (gauginos) are set to zero. Now it is very easy to see 
that the BPS condition (3.12) does imply the (Yang-Mills) equation of motion 
on the vector gauge field $A_{\m}^a$. The only remaining issue are the 
equations on the scalars $\bar{\f}^a$ and $\f^a$.

The equations of motion of $\f^a$, subject to the BPS equations (3.12) and 
(3.14), read
$$ \eqalign{
0 &~ = 4F'(\bar{\f}^2)\bar{\f}^a\bar{\f}^b\cd_{\m}\cd_{\m}\f^b 
+ 2F(\bar{\f}^2)\cd_{\m}\cd_{\m}\f^a \cr
&~ -F'(\bar{\f}^2)\bar{\f}^a\left(\f^2\bar{\f}^2-(\f\cdot\bar{\f})^2\right)
- F'(\bar{\f}^2)\left(\bar{\f}^a\f^2-\f^a\f\cdot\bar{\f}\right)~.\cr}
\eqno(3.17)$$
The equations of motion on $\bar{\f}^a$ take another form, 
$$ 2\cd_{\m}\cd^{\m}\left(F(\bar{\f}^2)\bar{\f}^a\right)-
F(\bar{\f}^2)\left(\f^a\bar{\f}^2-\bar{\f}^a\f\cdot\bar{\f}\right)=0~,
\eqno(3.18)$$
while they also follow from the Lagrangian (3.6). 

It is not difficult to verify that all scalar equations of motion (3.17) and 
(3.18) follow from the first-order equations
$$ \cd_{\m}\f^a=0\qquad {\rm and}\qquad \cd_{\m}
\left[F(\bar{\f}^2)\bar{\f}^a\right]=0~,\eqno(3.19)$$
subject to an (apparently consistent) algebraic constraint 
$$ \bar{\f}^a\f^2-\f^a(\f\cdot\bar{\f})=0~~.\eqno(3.20)$$
We propose eqs.~(3.19) and (3.20) as the BPS conditions supplementing 
eqs.~(3.12) and (3.14). We believe that those equations (in the presence of all
fermionic terms) preserve half of supersymmetry, though we didn't verify it.

\section{$U(2)$ NAC-deformed N=2 scalar potential}

Perhaps, the most interesting part of the NAC-deformed N=2 super-Yang-Mills 
Lagrangian is its scalar potential, because it is completely fixed by the 
choice of a NAC-deformation, a star product and a gauge group. It is worth
mentioning that no scalar potential is generated by a NAC deformation in the 
case of an abelian gauge group \cite{fs,ks1}. In the case of the simple 
$SU(2)$ gauge group, the NAC $(P)$ deformed N=2 scalar potential was 
calculated in ref.~\cite{ks2}, 
$$ V_{\rm NAC,~SU(2)} = -\fracmm{1}{4}F(\bar{\f}^2)
\Tr\[\f,\bar{\f}\]^2\equiv F(\bar{\f}^2)V_{\rm SYM}~,\eqno(4.1)$$
where we have explicitly introduced the undeformed (non-abelian) N=2 
super-Yang-Mills scalar potential $V_{\rm SYM}$. Equation (3.7) implies that
$$ V_{\rm NAC,~SU(2)} =\fracm{1}{2}F(\bar{\f}^2) \ve^{abf}\f^a\bar{\f}^b
\ve^{cdf}\f^c\bar{\f}^d  = 
\fracmm{(1-3\tilde{P}^2\bar{\f}^2)}{2(1+\tilde{P}^2\bar{\f}^2)^3}\left[
 \f^2\bar{\f}^2-(\f^a\bar{\f}^a)^2\right]~.\eqno(4.2)$$
When using the notation
$$ (\f^a\bar{\f}^a)^2= \f^2\bar{\f}^2\cos^2\vq~, \eqno(4.3)$$
equation (4.2) reads
$$  V_{\rm NAC,~SU(2)}(\f,\bar{\f})=\fracm{1}{2}
\f^2\bar{\f}^2\sin^2\vq\,\fracmm{1-3\tilde{P}^2\bar{\f}^2}{(1+\tilde{P}^2
\bar{\f}^2)^3}~~.\eqno(4.4)$$
The scalar potential $V_{\rm SYM}$ of the undeformed N=2 super-Yang-Mills 
theory is bounded from below (actually, non-negative), while the undeformed 
(and degenerate) classical vacua are given by solutions to the equation
$$ \[\f,\bar{\f}\]=0~.\eqno(4.5)$$
In the deformed $SU(2)$ case the fields $\f$ and $\bar{\f}$ are real and 
independent, while the $P$-deformation gives rise to the extra factor 
$F(\bar{\f}^2)$ in eqs.~(4.1) and (4.4). As was shown in ref.~\cite{ks2}, the
real scalar potential (4.1) is either singular (given a purely imaginary 
$\tilde{P}$) or unbounded from below (given a real $\tilde{P}>0$), so that the
classical  NAC deformed $SU(2)$-based N=2 supersymmetric gauge field 
theory does not have a stable vacuum. In this section we would like to 
calculate the scalar potential in the NAC theory with the non-simple 
$U(2)=SU(2)\times U(1)$ gauge group.

The $U(2)$-based action is given by eq.~(1.7), while the group-theoretical 
trace in eq.~(1.7) is given by the master function (2.13) of two colorless
variables $\Bar{W}{}^0$ and $\sqrt{\Bar{W}{}^a\Bar{W}{}^a}$. Being
N=2 superfields, those variables can be expanded in terms of the $N=2$ 
superspace Grassmann coordinates as
$$ \Bar{W}{}^0 = \bar{\f}^0 + \tilde{W}^0 \eqno(4.6)$$
and
$$\sqrt{\Bar{W}{}^a\Bar{W}{}^a}  = \sqrt{\Bar{\f}{}^a\Bar{\f}{}^a}
(1+\tilde{W})^{1/2}\equiv  \sqrt{\Bar{\f}{}^a\Bar{\f}{}^a}+ \hat{W}~~,
\eqno(4.7)$$
where $\tilde{W}^0$ and $\tilde{W}$ comprise all the Grassmann (nilpotent) 
terms of $\bar{W}{}^0$ and ${\Bar{W}{}^a\Bar{W}{}^a}$, respectively --- 
see eq.~(3.4) ---  and
$$ \hat{W}= \sqrt{\Bar{\f}{}^a\Bar{\f}{}^a}\left( \fracm{1}{2}\tilde{W} 
-\fracm{1}{8}\tilde{W}^2+\fracm{1}{16}\tilde{W}^3 -\fracm{5}{128}
\tilde{W}^4\right)~~.\eqno(4.8)$$
 
Similarly, any N=2 anti-chiral superfield function  
 $f\left(\sqrt{\Bar{W}{}^a\Bar{W}{}^a}\right)$ can be expanded with respect to
 the nilponent part of its argument as follows:
$$\eqalign{
f\left(\sqrt{\Bar{W}{}^a\Bar{W}{}^a}\right) 
&~ = f(\sqrt{\bar{\f}{}^a\bar{\f}{}^a}+\hat{W}) \cr
&~ =f(\sqrt{\bar{\f}{}^a\bar{\f}{}^a})+f'(\sqrt{\bar{\f}{}^a\bar{\f}{}^a})
\hat{W} + \fracmm{1}{2!}f''(\sqrt{\bar{\f}{}^a\bar{\f}{}^a})\hat{W}^2 \cr
&~ + \fracmm{1}{3!}f'''(\sqrt{\bar{\f}{}^a\bar{\f}{}^a})\hat{W}^3 +
\fracmm{1}{4!}f''''(\sqrt{\bar{\f}{}^a\bar{\f}{}^a})\hat{W}^4~~,\cr}
\eqno(4.9)$$
and similarly for any function of $\Bar{W}^0=\bar{\f}^0+\tilde{W}^0$.

It is now fully straightfoward to perform the Grassmann integration over 
$d^4\bar{\q}$ in eq.~(1.7). The scalar potential is generated from the second
term in eq.~(4.9) by using the master function (2.13). We find
$$ V_{\rm NAC,~U(2)}= \fracmm{1}{4}
\fracmm{\pa h(\bar{\f}^0,\sqrt{\bar{\f}{}^a\bar{\f}{}^a})}{\pa
(\sqrt{\bar{\f}{}^a\bar{\f}{}^a})}\sqrt{\bar{\f}{}^a\bar{\f}{}^a}
(\f^b\f^b)\sin^2\vq~~,\eqno(4.10)$$
or more explicitly,
$$  V_{\rm NAC,~U(2)}=\fracmm{1-3P^2(\bar{\f}^0\bar{\f}^0
-\bar{\f}^a\bar{\f}^a)-2P^3\bar{\f}^0\bar{\f}^0(\bar{\f}^0\bar{\f}^0
-\bar{\f}^a\bar{\f}^a)}{\left[ 1+P(\bar{\f}^0-\sqrt{\bar{\f}{}^a\bar{\f}{}^a})
\right]^3\left[ 1+P(\bar{\f}^0+\sqrt{\bar{\f}{}^a\bar{\f}{}^a})\right]^3}
\bar{\f}^a\bar{\f}^a(\f^b\f^b)\sin^2\vq~.\eqno(4.11)$$
In terms of the new variables $x=\f^0$, $y=\sqrt{\bar{\f}{}^a\bar{\f}{}^a}
\geq 0$ and $z^2=\f^a\f^a\geq 0$, the potential reads
$$ V_{\rm NAC,~U(2)}=\fracmm{1-3P^2(x^2-y^2)-2P^3x(x^2-y^2)}{(1+Px-Py)^3
(1+Px+Py)^3}y^2(z^2\sin^2\vq)~.\eqno(4.12)$$ 
A graph of the potential (4.12) in $(x,y)$ variables is given by Fig.~1 
(the factorized $(z,\vq)$-dependence is trivial). 

\begin{figure}[t]
\begin{center}
\vglue.1in
\makebox{
\epsfxsize=3in
\epsfbox{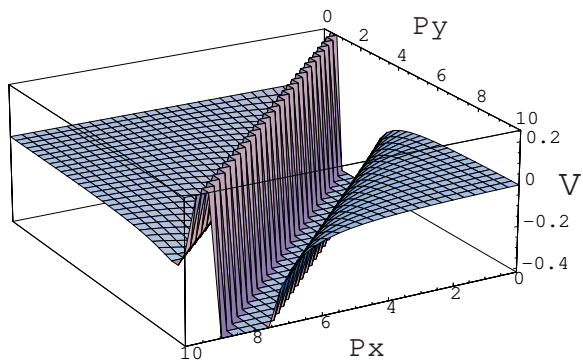}
  }
\caption{\small Graph of the potential}
\end{center}
\end{figure}

The scalar potential (4.12) reduces to the standard (bounded from below) 
scalar potential $V\propto\Tr\[\f,\bar{\f}\]^2$ of the undeformed N=2 
super-Yang-Mils theory with the $U(2)$ gauge group in the limit $P\to 0$. 
Equation (4.12) vanishes in the NAC-deformed $U(1)$ gauge group case, when 
$y=0$, as it should have been expected. Finally, the scalar potential (4.12) 
reduces to that of eq.~(4.4) in the $SU(2)$ limit $x=0$. 

The $U(2)$-based potential (4.12) suffers from singularities while it is not
positively definite, so that it also implies the absence of a well-defined 
vacuum in the classical NAC, $U(2)$-based N=2 supersymmetric super-Yang-Mills 
theory.

\section{Conclusion}

Our investigation of the NAC-deformed N=2 supersymmetric gauge field theory 
with a non-simple gauge group $U(2)$ (i.e. with two gauge coupling constants)  
reveals the rich structure of the effective theory (1.7) after performing the
Seiberg-Witen map and the group-theoretical trace. We were able to explicitly
calculate the master function governing the component structure of the theory
under consideration, as well as its highly non-trivial scalar potential. It 
should be emphasized that our results are truly non-perturbative in the sense 
that they include all corrections in the NAC-deformation parameter $P$.

The case of the $U(2)$ gauge group also clearly illustrates the difficulties
arising in the efforts to generalize our results to the other gauge groups
different from $SU(2)$ and $U(1)$. For instance, in the case of the $SU(3)$ 
gauge group, the master function is going to depend upon two variables 
$$ \Bar{W}{}^a\Bar{W}{}^a\equiv \d^{ab}\Bar{W}{}^a\Bar{W}{}^b\quad {\rm  and}
\quad d^{abc}\Bar{W}{}^a\Bar{W}{}^b\Bar{W}{}^c~~,\eqno(5.1)$$
in terms of the two $SU(3)$-invariant symmetric tensors $\d^{ab}$ and
$d^{abc}$, where $a,b,c=1,2,\ldots,8$. By using the structure constants of
$SU(3)$ we were able to calculate the leading NAC-contribution to the
$SU(3)$  master function,
$$ \Tr f(\Bar{W}{}^aT^a) = 3(\Bar{W}{}^a\Bar{W}{}^a)  -
\fracmm{27P^2}{4}(\Bar{W}{}^a\Bar{W}{}^a)^2 +{\cal O}(P^4)~,\eqno(5.2)$$ 
but we faced considerable technical difficulties in calculating the next order
 terms, not to mention a full non-perturbative answer.

Our considerations in this paper were entirely classical. It is conceivable,
however, that the NAC-deformed supersymmetric gauge field theories may even be 
renormalizable in some sense. So it would be interesting to investigate the 
role of quantum corrections to eq.~(1.7), both in quantum field theory and in 
superstring theory (e.g., by using geometrical engineering). It is 
particularly intriguing to know whether quantum corrections can stabilize the 
classical vacuum.


\begin{thebibliography}{99}

\bibitem{rev} M. R. Douglas and N. A. Nekrasov, Rev. Mod. Phys. {\bf 73} (2001)
[hep-th/0106048]
\bibitem{li} M. Li and Y. S. Wu, Editors, 
{\it Physics in Noncommutative World}, Rinton Press, New Yersey, 2002
\bibitem{sw} N. Seiberg and E. Witten, JHEP {\bf 9909} (1999) 032
[hep-th/9908142];\\
H. Liu, Nucl. Phys. {\bf B614} (2001) 305 [hep-th/0011125];\\
Y. Okawa and H. Ooguri, Phys. Rev. {\bf D64} (2001) 046009 [hep-th/0104036];\\
Ch. Saemann and M. Wolf, JHEP {\bf 0403} (2004) 048  [hep-th/0401147];\\
D. Mikulovic, JHEP {\bf 0405} (2004) 077 [hep-th/0403290]
\bibitem{sei} N. Seiberg, JHEP {\bf 0306} (2003) 010 [hep-th/0305248]
\bibitem{nac} L. Brink and J. H. Schwarz, Phys. Lett. {\bf B100} (1981) 310;\\
J. H. Schwarz and P. van Nieuwenhuizen, Lett. Nuovo Cim. {\bf 34} (1982) 21;\\
S. Ferrara and M. A. Lledo, JHEP {\bf 0005} (2000) 008 [hep-th/0002084];\\
D. Klemm, S. Penati and L. Tamassia, Class. Quantum Grav. {\bf 20} (2003) 2905 
[hep-th/0104190]
\bibitem{rec} J. de Boer, P. A. Grassi and P. van Nieuwenhuizen, Phys.Lett. 
{\bf B574} (2003) 98 [hep-th/0302078];\\
A. Imaanpur, JHEP {\bf 0309} (2003) 077 [hep-th/0308171];\\
R. Britto, B. Feng, O. Lunin, S.-J. Rey, JHEP {\bf 0307} (2003) 067 
 [hep-th/0311275]
\bibitem{dv} R. Dijkgraaf and C. Vafa, Nucl. Phys. {\bf B644} (2002) 3 and 21;
[hep-th/0206255 and 0207106]; and  hep-th/0208048
\bibitem{ovn} H. Ooguri and C. Vafa, Adv. Theor. Math. Phys. {\bf 7} (2003) 
53 and {\bf 7} (2004) 405;  [hep-th/0302109 and 0303063]  
\bibitem{ll} M. Billo, M. Frau, I. Pesando and A. Lerda, JHEP {\bf 0405} 
(2004) 023 [hep-th/0402160]
\bibitem{ov} H. Ooguri and C. Vafa, Mod. Phys. Lett. {\bf A5} (1990) 1389;
Nucl. Phys. {\bf B361} (1991) 469 {\it ibid.} {\bf B367} (1991) 83;\\
J. Bischoff, S. V. Ketov and O. Lechtenfeld, Nucl.Phys. {\bf B438}
(1995) 373 \newline[hep-th/9406101] 
\bibitem{mary} S. J. Gates, Jr., S. V. Ketov and H. Nishino,
Phys.Lett. {\bf B297} (1992) 99 [hep-th/9203078]; {\it ibid.} {\bf B303} 
(1993) 323 [hep-th/9203081]; Nucl. Phys. {\bf B393} (1993) 149 [hep-th/9207042]
\bibitem{2nac} T. Araki, L. Ito and A. Ohtsuka, Phys.Lett. {\bf B573} (2003) 
209 [hep-th/0307076];\\
E. Ivanov, O. Lechtenfeld, B. Zupnik, JHEP {\bf 0402} (2004) 012 
[hep-th/0308012];\\
T. Araki and K. Ito, hep-th/0404250;\\
S. Ferrara, E. Ivanov, O. Lechtenfeld, E. Sokatchev, B. Zupnik, hep-th/0405049
\bibitem{fs} S. Ferrara and E. Sokatchev, Phys.Lett. {\bf B579} (2004) 226 
[hep-th/0308021]
\bibitem{book} S. V. Ketov, {\it Quantum Non-linear Sigma-models}, 
Springer-Verlag, 2000, Sect.~4.2
\bibitem{ks1} S. V. Ketov and S. Sasaki, Phys. Lett. {\bf B595} (2004) 530
 [hep-th/0404119] 
\bibitem{ks2} S. V. Ketov and S. Sasaki, hep-th/0405278; to appear in Phys. 
Lett. B.
\end{thebibliography}
\end{document}